\documentclass[runningheads]{llncs}
\usepackage[T1]{fontenc}
\usepackage{graphicx}
\usepackage[colorlinks=true, allcolors=blue]{hyperref}
\usepackage{color}

\usepackage{multirow}
\usepackage{amsmath}
\usepackage{siunitx}
\usepackage{booktabs}
\begin{document}
\title{TAI-GAN: Temporally and Anatomically Informed GAN for early-to-late frame conversion in dynamic cardiac PET motion correction}
\titlerunning{TAI-GAN frame conversion in dynamic PET MC}
%
\author{Xueqi Guo\inst{1}\orcidID{0000-0002-0416-2811} \and Luyao Shi\inst{2}\orcidID{0000-0002-4703-3294} \and Xiongchao Chen\inst{1}\orcidID{0000-0003-4112-8492} \and Bo Zhou\inst{1}\orcidID{0000-0002-2906-0897} \and Qiong Liu\inst{1}\orcidID{0009-0003-9420-8549} \and Huidong Xie\inst{1}\orcidID{0000-0002-1124-3548} \and Yi-Hwa Liu\inst{1} \and Richard Palyo\inst{3} \and Edward J. Miller\inst{1}\orcidID{0000-0002-2156-5962} \and Albert J. Sinusas\inst{1}\orcidID{0000-0003-0972-9589} \and Bruce Spottiswoode\inst{4} \and Chi Liu\inst{1} \and Nicha C. Dvornek\inst{1}\orcidID{0000-0002-1648-6055}}
\authorrunning{Guo et al.}
%
\institute{Yale University, New Haven CT 06511, USA \\
\email{\{xueqi.guo,chi.liu,nicha.dvornek\}@yale.edu} \and
IBM Research, San Jose CA 95120, USA \and
Yale New Haven Hospital, New Haven CT 06511, USA \and
Siemens Medical Solutions USA, Inc., Knoxville TN 37932, USA}
\maketitle              
\begin{abstract}
The rapid tracer kinetics of rubidium-82 ($^{82}$Rb) and high variation of cross-frame distribution in dynamic cardiac positron emission tomography (PET) raise significant challenges for inter-frame motion correction, particularly for the early frames where conventional inten-sity-based image registration techniques are not applicable. Alternatively, a promising approach utilizes generative methods to handle the tracer distribution changes to assist existing registration methods. To improve frame-wise registration and parametric quantification, we propose a Temporally and Anatomically Informed Generative Adversarial Network (TAI-GAN) to transform the early frames into the late reference frame using an all-to-one mapping. Specifically, a feature-wise linear modulation layer encodes channel-wise parameters generated from temporal tracer kinetics information, and rough cardiac segmentations with local shifts serve as the anatomical information. We validated our proposed method on a clinical $^{82}$Rb PET dataset and found that our TAI-GAN can produce converted early frames with high image quality, comparable to the real reference frames. After TAI-GAN conversion, motion estimation accuracy and clinical myocardial blood flow (MBF) quantification were improved compared to using the original frames. Our code is published at \href{https://github.com/gxq1998/TAI-GAN}{https://github.com/gxq1998/TAI-GAN}.

\end{abstract}
\section{Introduction}
Compared to other non-invasive imaging techniques, dynamic cardiac positron emission tomography (PET) myocardial perfusion imaging increases the accuracy of coronary artery disease detection \cite{prior2012quantification}. After tracer injection, a dynamic frame sequence is acquired over several minutes until the myocardium is well perfused. The time-activity curves (TACs) are collected in the myocardium tissue and left ventricle blood pool (LVBP) using regions of interest (ROIs) derived from the reconstructed frames. The myocardial blood flow (MBF) is then quantified through kinetic modeling using myocardium and LVBP TACs.

However, the inter-frame motion will cause spatial misalignment across the dynamic frames, resulting in incorrect TAC measurements and a major impact on both ROI-based and voxel-based MBF quantification \cite{hunter2016patient}. The high variation of cross-tracer distribution originating from the rapid tracer kinetics of rubidium-82 ($^{82}$Rb) further complicates inter-frame motion correction, especially for the early frames when the injected tracer is concentrated in the blood pool and has not been well distributed in the myocardium. Most existing motion correction studies and clinical software focus solely on the later frames in the myocardial perfusion phase \cite{woo2011automatic,lu2020patient,burckhardt2009cardiac}. Although deep learning-based dynamic PET motion correction has outperformed conventional techniques \cite{guo2021interframe,guo2022mcp,zhou2023fast}, few focused on $^{82}$Rb cardiac PET due to the higher difficulty. An automatic motion correction network was proposed for $^{82}$Rb cardiac PET under supervised learning using simulated translational motion \cite{shi2021automatic}, but the method requires training two separate models to handle the discrepancy between early and late frames, which is inconvenient and computationally expensive. 

Alternatively, the usage of image synthesis and modality conversion has been proposed to improve optimization in multi-modality image registration \cite{cao2018region,liu2019image,maul2021x}, mostly involving magnetic resonance imaging. In FDG dynamic PET, converting early frames to the corresponding late frame using a generative adversarial network (GAN) is a promising way to overcome the barrier of tracer differences and aid motion correction \cite{sundar2021conditional,sundar2021data}. However, this method trains one-to-one mappings for each specific early frame, which is impractical to implement and also difficult to generalize to new acquisitions. Moreover, the tracer kinetics and related temporal analysis are not incorporated in model training, which might be a challenge when directly applied to $^{82}$Rb cardiac PET.

In this work, we propose a Temporally and Anatomically Informed GAN (TAI-GAN) as an all-to-one mapping to convert all the early frames into the last reference frame. A feature-wise linear modulation (FiLM) layer encodes channel-wise parameters generated from the blood pool TACs and the temporal frame index, providing additional temporal information to the generator. The rough segmentations of the right ventricle blood pool (RVBP), LVBP, and myocardium with local shifts serve as the auxiliary anatomical information. Most current work applying GAN+FiLM models encode text or semantic information for natural images \cite{ak2019semantically,mao2019bilinear,ak2020semantically} and metadata for medical images \cite{rachmadi2019predicting,dey2021generative}, while we innovatively propose encoding dynamic PET tracer distribution changes. TAI-GAN is the first work incorporating both temporal and anatomical information into the GAN for dynamic cardiac PET frame conversion, with the ability to handle high tracer distribution variability and prevent spatial mismatch. 

\section{Methods}
\subsection{Dataset}
This study includes 85 clinical $^{82}$Rb PET scans (55 rest and 30 regadenoson-induced stress) that were acquired from 59 patients at the Anonymous Institution using a GE Discovery 690 PET/CT scanner and defined by clinicians to be nearly motion-free, with Institutional Review Board approval. After weight-based $^{82}$Rb injection, the list-mode data of each scan for the first 6 minutes and 10 seconds were rebinned into 27 dynamic frames (14×5s, 6×10s, 3×20s, 3×30s, 1×90s). The details of the imaging and reconstruction protocol are in Supplementary Figure S1. In all the scans, the rough segmentations of RVBP, LVBP, and myocardium were manually labeled with reference to the last dynamic frame for TAC generation and the following MBF quantification. 

\subsection{Network architecture}

\begin{figure}[t]
\centering
\includegraphics[width=0.85\textwidth,keepaspectratio]{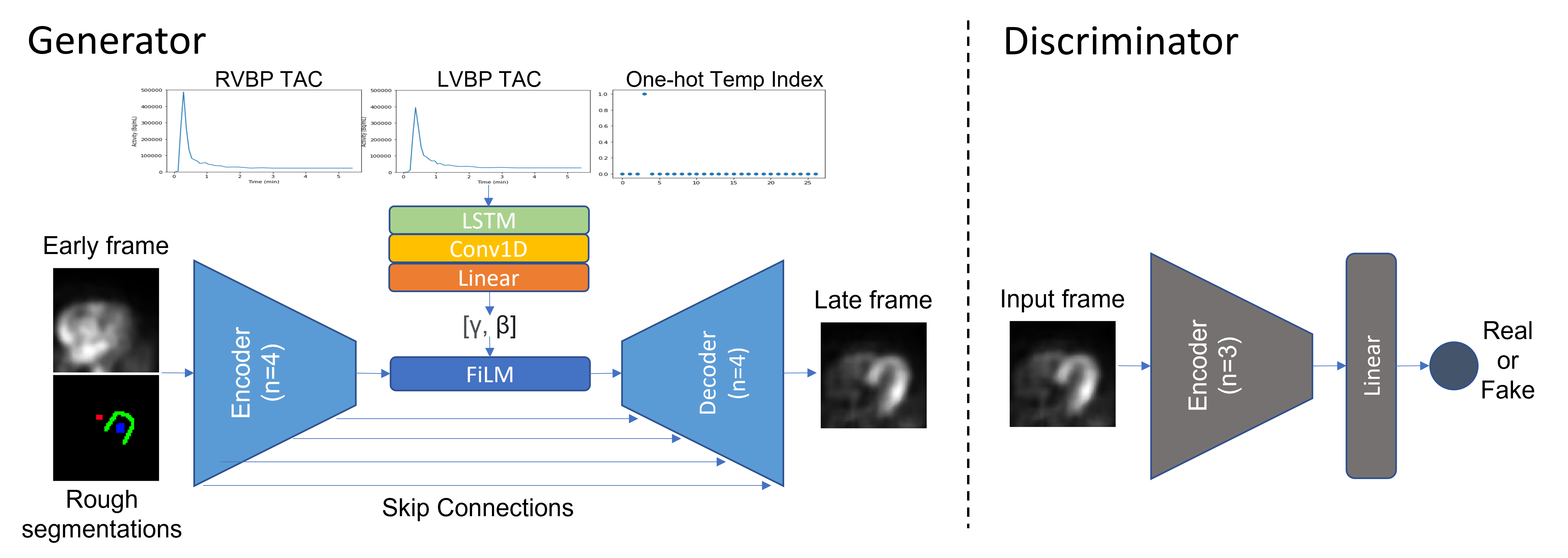}
\caption{The structure of the proposed early-to-late frame conversion network TAI-GAN.} 
\label{network}
\end{figure}

The structure of the proposed network TAI-GAN is shown in Figure \ref{network}. The generator predicts the related late frame using the input early frame, with the backbone structure of a 3-D U-Net \cite{cciccek20163d} (4 encoding and decoding levels), modified to be temporally and anatomically informed. The discriminator analyzes the true and generated late frames and categorizes them as either real or fake, employing the structure of PatchGAN \cite{isola2017image} (3 encoding levels, 1 linear output layer).

\subsubsection{Temporally informed by tracer dynamics and FiLM} To address the high variation in tracer distribution in the different phases, the temporal information related to tracer dynamics is introduced to the network by concatenating RVBP and LVBP TACs as well as the frame temporal index in one-hot format. A long short-term memory (LSTM) \cite{hochreiter1997long} layer encodes the concatenated temporal input, and the following 1-D convolutional layer and linear layer map the LSTM outputs to the channel-wise parameters $\gamma$ and $\beta$. The feature-wise linear modulation (FiLM) \cite{perez2018film} layer then manipulates the bottleneck feature map by the generated scaling factor $\gamma$ and bias $\beta$, as in \eqref{FiLM},
\begin{equation}
\label{FiLM}
FiLM(M_i) = \gamma_i \cdot M_i+\beta_i,
\end{equation}
where $M_i$ is the $i^{th}$ channel of the bottleneck feature map, 
$\gamma_i$ and $\beta_i$ are the scaling factor and the bias of the $i^{th}$ channel, respectively.

\subsubsection{Anatomically informed by segmentation locators} 
The dual-channel input of the generator is the early frame concatenated with the rough segmentations of RVBP, LVBP, and myocardium. Note that cardiac segmentations are already required in MBF quantification and this is an essential part of the current clinical workflow. In our work, the labeled masks serve as the anatomical locator to inform the generator of the cardiac ROI location and prevent spatial mismatch in frame conversion. This is especially helpful in discriminating against the frame conversion of early RV and LV phases. Random local shifts of the segmentations are applied during training. This improves the robustness of the conversion network to motion between the early frame and the last frame. 

\subsubsection{Loss function} 
Both an adversarial loss and a voxel-wise mean squared error (MSE) loss are included in the loss function of TAI-GAN, as in \eqref{loss_1}-\eqref{loss_3}, 
\begin{align}
&L_{adv}= -log(D(F_L)-log(1-D(G(F_i))) \label{loss_1}\\
&L_{mse}= \frac{1}{V}\sum\limits_{n=1}^{V}(G(F_i)_n - (F_L)_n)^2\label{loss_2}\\
&\hat{G},\hat{D}= arg\min_{G}\max_{D}(L_{adv}+L_{mse})\label{loss_3}
\end{align}
where $L_{adv}$ is the adversarial loss, $L_{mse}$ is the MSE loss, $D$ is the discriminator, $G$ is the generator, $F_L$ is the real last frame, $G(F_i)$ is the generator-mapped last frame from the $i^{th}$ early frame $F_i$, and $V$ is the number of voxels in each frame. 

\subsection{Network training and image conversion evaluation}
All the early frames with LVBP activity higher than 10\% of the maximum activity in TAC are converted to the last frame. Very early frames with lower activity in LVBP are not considered as they do not have a meaningful impact on the image-derived input function and subsequently the associated MBF quantification \cite{shi2021automatic}. Prior to model input, all the frames were individually normalized to the intensity range of [-1,1]. Patch-based training was implemented with random cropping size of (64,64,32) near the location of LV inferior wall center, random rotation in xy plane in the range of [-45$^{\circ}$,45$^{\circ}$], and 3-D random translation in the range of [-5,5] voxels as the data augmentation.

Considering the low feasibility of training each one-to-one mapping for all the early frames, we trained two pairwise mappings using a vanilla GAN (3-D U-Net generator) and solely the adversarial loss as a comparison with \cite{sundar2021conditional}. The two specific mappings are one frame before and one frame after the EQ frame, the first frame when LVBP activity is equal to or higher than RVBP activity \cite{shi2021automatic}, respectively EQ-1 and EQ+1. We also implemented the vanilla GAN and the MSE loss GAN as two all-to-one conversion baselines. A preliminary ablation study of the introduced temporal and anatomic information is summarized in Supplementary Figure S2 and Table S1. A comparison of the average training time and memory footprint is included in Supplementary Table S2. 

All the deep learning models are developed using PyTorch and trained under 5-fold cross-validation on an NVIDIA A40 GPU using Adam optimizer (learning rate $G$=2e-4, $D$=5e-5). In each fold, 17 scans were randomly selected as the test set and the remaining 68 were for training. The stopping epoch was 800 for one-to-one mappings and 100 for all the all-to-one models.

Image conversion evaluations include visualizing the generated last frames against the real last frame and the input frame, with overlaid cardiac segmentations. Quantitatively, the MSE, normalized mean absolute error (NMAE), peak signal-to-noise ratio (PSNR), and structural similarity index (SSIM) are computed between the generated and real last frames. Differences between methods were assessed by fold-wise paired two-tailed t-tests ($\alpha$ = 0.05).

\begin{figure}[t]
\centering
\includegraphics[width=0.7\textwidth,keepaspectratio]{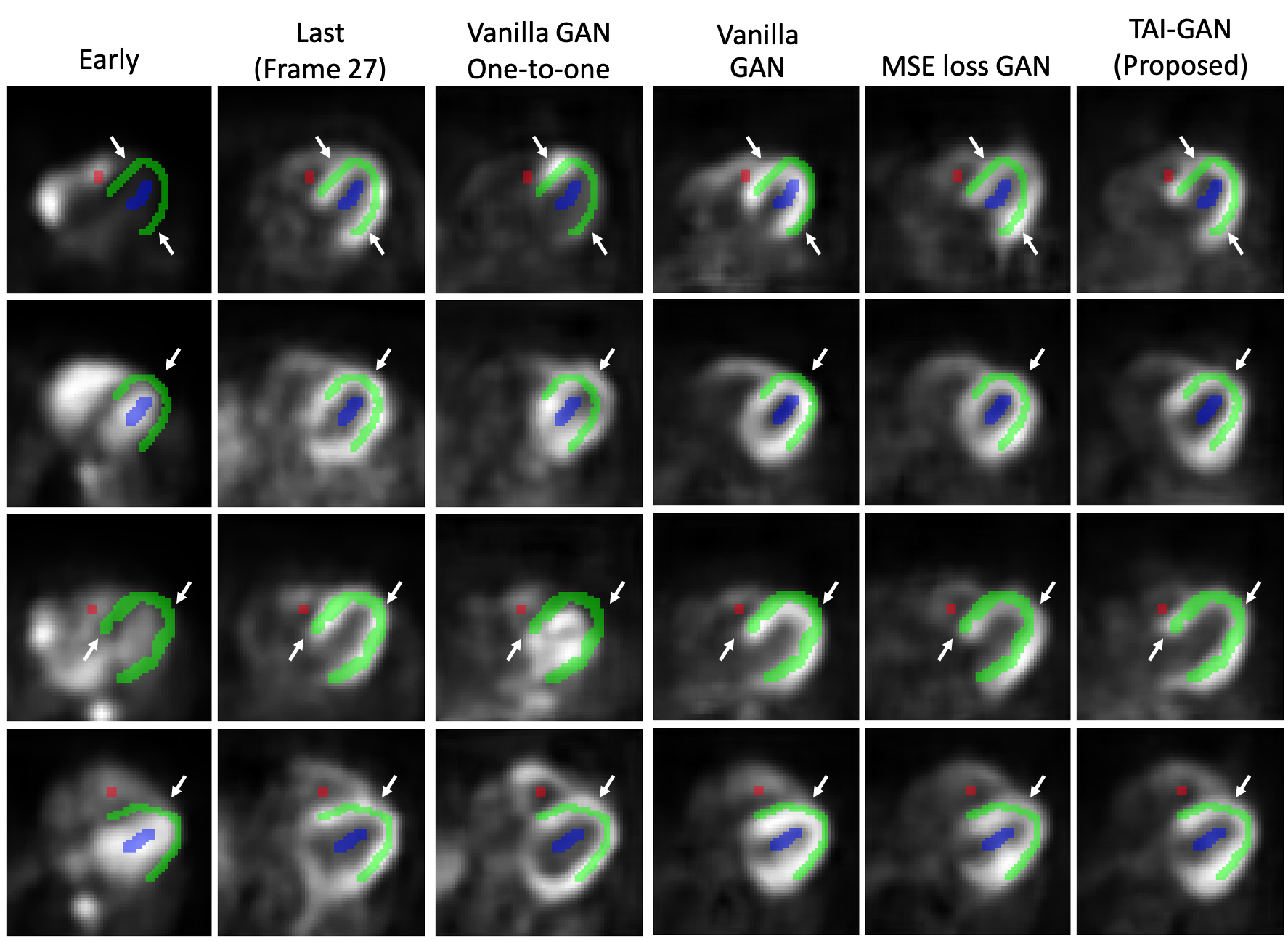}
\caption{Sample early-to-late frame conversion results of each method with overlaid segmentations of RVBP (red), LVBP (blue), and myocardium (green).}
\label{conversion_result}
\end{figure}

\subsection{Motion correction and clinical MBF quantification}
Since all the included scans are categorized as motion-free, we ran a motion simulation test to evaluate the benefit of frame conversion on motion correction using the test set of one random fold, resulting in 17 cases. On an independent $^{82}$Rb cardiac scan cohort identified as having significant motion by a clinician, we ran non-rigid motion correction in BioImage Suite \cite{joshi2011unified} (BIS) to generate motion fields. We applied the motion field estimations from the late frames scaled by 2 to the motion-free test frames as motion ground-truth. In this way, the characteristics of simulated motion match with real-patient motion and also have significant magnitudes. The different image conversion methods were applied to the early frames prior to motion simulation. All original and converted frames with simulated motion were then registered to the last frame using BIS with the settings as in \cite{guo2022inter}. 
We calculated the mean absolute prediction error to measure motion prediction accuracy, 
\begin{equation}
\label{motion_error}
|\Delta \Phi| = \frac{1}{P}\sum\limits_{n=1}^{P}\frac{|\Phi_{x_n} - \hat{\Phi}_{x_n}| + |\Phi_{y_n} - \hat{\Phi}_{y_n}| + |\Phi_{z_n} - \hat{\Phi}_{z_n}|}{3},
\end{equation}
where $P$ is the number of transformation control points in a frame, ($\hat{\Phi}_{x_n}$,$\hat{\Phi}_{y_n}$,$\hat{\Phi}_{z_n}$) is the motion prediction, and ($\Phi_{x_n}$,$\Phi_{y_n}$,$\Phi_{z_n}$) is the motion ground-truth. 

After motion estimation, the predicted motion of each method is applied to the original frames without intensity normalization for kinetic modeling. To estimate the uptake rate $K_1$, the LVBP TAC as the image-derived input function and myocardium TAC were fitted to a 1-tissue compartment model using weighted least squares fitting as in \cite{shi2019direct}. MBF was then calculated from $K_1$ under the relationship as in \cite{germino2016quantification}. The percentage differences of $K_1$ and MBF were calculated between the motion-free ground-truth and motion-corrected values. The weighted sum-of-squared (WSS) residuals were computed between the MBF model predictions and the observed TACs. We also included a comparison of LVBP and myocardium TACs in Supplementary Figure S3. 

\newcommand{\tabincell}[2]{\begin{tabular}{@{}#1@{}}#2\end{tabular}}
\begin{table}[t]
\centering
\sisetup{
    detect-weight=true,
    mode=text,
  }
\caption{Quantitative image similarity evaluation of early-to-late frame conversion (mean $\pm$ standard deviation) with the best results marked \textbf{in bold}.}
\label{tab1}
\resizebox{0.7\textwidth}{!}{
\begin{tabular}{c|c|S[table-format=2.3(3)] c |S[table-format=2.3(3)] c|S[table-format=2.3(3)] c|S[table-format=2.3(3)]}
\hline
Test set & Metric & {\tabincell{c}{Vanilla GAN\\One-to-one}} && {\tabincell{c}{Vanilla \\GAN}} && {\tabincell{c}{MSE loss \\GAN}} && {\tabincell{c}{TAI-GAN\\(Proposed)}}\\
\hline
\multirow{4}{*}{EQ-1} & SSIM & 0.557 \pm 0.017$^{*}$ & & 0.640 \pm 0.021 & &0.633 \pm 0.053$^{*}$ && \bfseries 0.657 \pm 0.018\\
 \cline{2-9}
 & MSE & 0.057 \pm 0.001$^{*}$ & & 0.050 \pm 0.006$^{*}$ && 0.044 \pm 0.011 & &\bfseries 0.040 \pm 0.005\\
 \cline{2-9}
 & NMAE & 0.068 \pm 0.002$^{*}$ & & 0.063 \pm 0.005$^{*}$ & &0.059 \pm 0.009 && \bfseries 0.057 \pm 0.005\\
 \cline{2-9}
 & PSNR & 18.678 \pm 0.116$^{*}$ & & 19.370 \pm 0.474$^{*}$ & &19.950 \pm 0.949 && \bfseries 20.335 \pm 0.530\\
 \hline
 \hline
\multirow{4}{*}{EQ+1} & SSIM & 0.669 \pm 0.061$^{*}$ & & 0.679 \pm 0.014 && 0.680 \pm 0.011 && \bfseries 0.691 \pm 0.013\\
 \cline{2-9}
 & MSE & 0.032 \pm 0.014 & & 0.034 \pm 0.002 & &0.033 \pm 0.006 & &\bfseries 0.032 \pm 0.002\\
 \cline{2-9}
 & NMAE & 0.050 \pm 0.011 & & 0.053 \pm 0.003 && 0.051 \pm 0.006 & &\bfseries 0.048 \pm 0.003\\
 \cline{2-9}
 & PSNR & 21.323 \pm 1.800 & & 21.014 \pm 0.355 & &21.188 \pm 0.757 & &\bfseries 21.361 \pm 0.205\\
 \hline
  \hline
\multirow{4}{*}{\tabincell{c}{All\\Pre-EQ\\frames}} & SSIM & {-}  & & 0.594 \pm 0.012$^{*}$ & &0.596 \pm 0.047$^{*}$ & &\bfseries 0.627 \pm 0.025\\
 \cline{2-9}
 & MSE & {-} & & 0.063 \pm 0.010$^{*}$ && 0.053 \pm 0.016 & & \bfseries 0.046 \pm 0.009\\
 \cline{2-9}
 & NMAE & {-} & & 0.072 \pm 0.007$^{*}$ & & 0.066 \pm 0.011 & & \bfseries 0.062 \pm 0.008\\
 \cline{2-9}
 & PSNR & {-} & & 18.507 \pm 0.474$^{*}$ & & 19.269 \pm 1.036$^{*}$ & & \bfseries 19.834 \pm 0.738\\
 \hline
  \hline

 \multirow{4}{*}{\tabincell{c}{All \\frames}} & SSIM & {-} & & 0.708 \pm 0.010$^{*}$ & & 0.716 \pm 0.007$^{*}$ & & \bfseries0.733 \pm 0.018\\

 \cline{2-9}
 & MSE & {-} & & 0.027 \pm 0.002$^{*}$ & & 0.024 \pm 0.003 & & \bfseries 0.021 \pm 0.002\\
 \cline{2-9}
 & NMAE & {-} & & 0.047 \pm 0.004$^{*}$ && 0.044 \pm 0.002 & &\bfseries0.040 \pm 0.002\\
 \cline{2-9}
 & PSNR & {-} & & 22.803 \pm 0.530$^{*}$ && 23.241 \pm 0.342$^{*}$ && \bfseries 23.799 \pm 0.466\\
 \hline
\multicolumn{9}{l}{{\footnotesize  $^{*}P < 0.05$ between the current method and TAI-GAN (paired two-tailed t-test).}}\\ 
\end{tabular}
}
\end{table}

\section{Results}
\subsection{Frame conversion performance}

The sample results of the early-to-late frame conversion by each method are visualized in Figure \ref{conversion_result}. Although the one-to-one models were trained under the most specific temporal mapping, the prediction results were not satisfactory and showed some failure cases, possibly due to the small sample size and the insufficient kinetics information of one early frame. Of the all-to-one models, vanilla GAN was able to learn the conversion patterns but with some distortions. After introducing MSE loss, the GAN generated results with higher visual similarity. After introducing temporal and anatomical information, the visual performance of the proposed TAI-GAN was the best with less mismatch and distortion.

The image similarity evaluation results are summarized in Table \ref{tab1}. The one-to-one training pair did not achieve better results than the all-to-one models, possibly due to the lack of inter-frame tracer dynamic dependencies. The TAI-GAN achieved the best result in each metric on each test set. The major improvement of the proposed TAI-GAN was for the Pre-EQ frames where the LVBP activity < RVBP activity, which is the most challenging to convert to late frame due to the large difference between the input and output frames. 

\subsection{Motion correction evaluation}
\begin{figure}[t]
\centering
\includegraphics[width=0.82\textwidth,keepaspectratio]{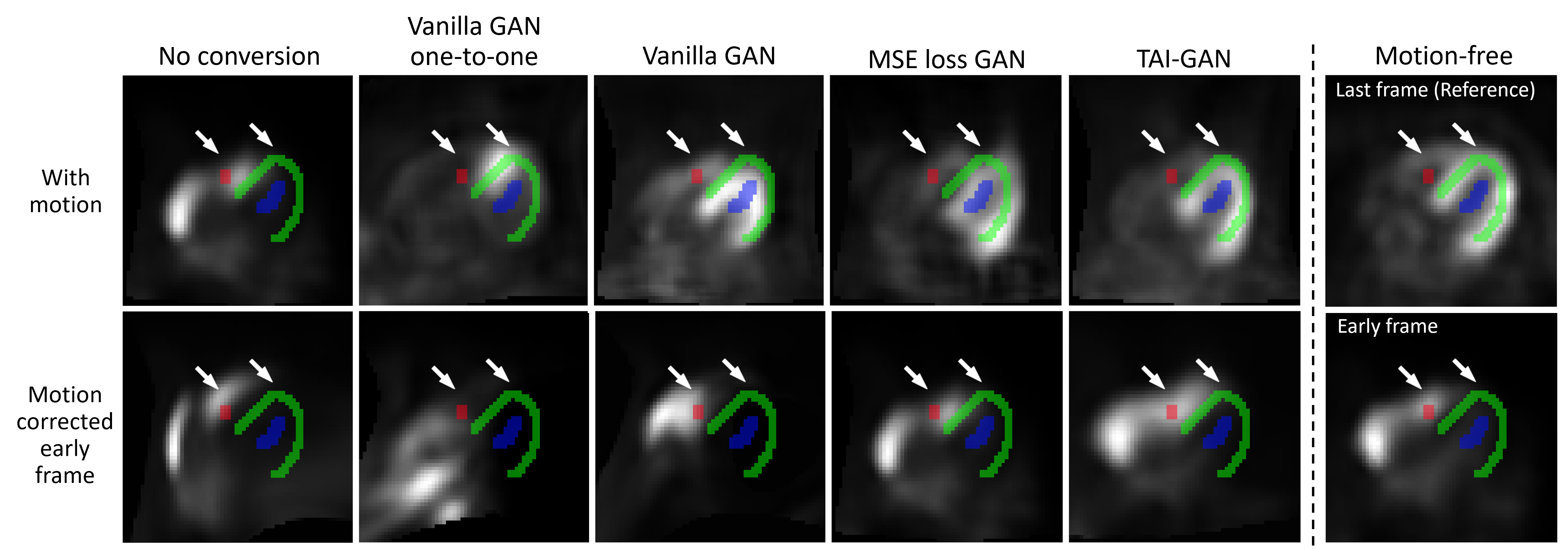}
\caption{Sample motion simulation and correction results with different methods of frame conversion.}
\label{motion_sim}
\end{figure}

Sample motion simulation and correction results are shown in Figure \ref{motion_sim}. The simulated non-rigid motion introduced distortion to the frames and the mismatch between the motion-affected early frame and the segmentation is observed. After directly registering the original frames, the resliced frame was even more deformed, likely due to the tracer distribution differences in the registration pair. Early-to-late frame conversion could address such challenging registration cases, but additional mismatches might be introduced due to conversion errors, as seen in the vanilla and MSE loss GAN results. With minimal local distortion and the highest frame similarity, the conversion result of the proposed TAI-GAN matched the myocardium and ventricle locations with the original early frame and the registration result demonstrated the best visual alignment.

\begin{table}[t]
\centering
\caption{Mean absolute motion prediction errors without and with each conversion method (in mm, mean ± standard deviation) with the best results marked \textbf{in bold}.}
\label{tab2}
\resizebox{0.67\textwidth}{!}{
\begin{tabular}{c|c|c|c|c|c}
\hline
& \tabincell{c}{No\\Conversion}& \tabincell{c}{Vanilla GAN\\One-to-one} & Vanilla GAN & \tabincell{c}{MSE loss \\GAN } & TAI-GAN\\
\hline
All frames & 4.45 ± 0.64$^{*}$ & - & 4.40 ± 0.49$^{*}$ & 4.76 ± 0.48$^{*}$ & \textbf{3.48 ± 0.45}\\
\hline
\tabincell{c}{EQ-1} & 6.18 ± 1.51$^{*}$ & 5.33 ± 1.34 & 6.03 ± 1.06$^{*}$ & 5.12 ± 0.72$^{*}$ & \textbf{5.06 ± 0.78}\\
\hline\tabincell{c}{EQ+1} & 5.12 ± 0.93$^{*}$ & 4.72 ± 0.86 & 4.93 ± 0.80$^{*}$ & 4.81 ± 0.46$^{*}$ & \textbf{4.35 ± 0.87}\\
\hline
\multicolumn{6}{l}{$^{*}$P < 0.05 between the current method and TAI-GAN (paired two-tailed t-test).}
\end{tabular}
}
\end{table}

Table \ref{tab2} summarizes the mean absolute motion prediction error on the original early frames and converted frames. Generally, the early acquisition time of a frame relates to high motion prediction errors. In all the included early frames and both EQ-1 and EQ+1 frames, the proposed TAI-GAN achieved the lowest motion prediction error and significantly reduced average prediction error compared to no conversion and all-to-one GAN models (p<0.05). The improvement of motion correction accuracy after proper frame conversion is suggested.

 \subsection{Parametric fitting and clinical MBF quantification}
 
\begin{figure}[t]
\centering
\includegraphics[width=0.9\textwidth,keepaspectratio]{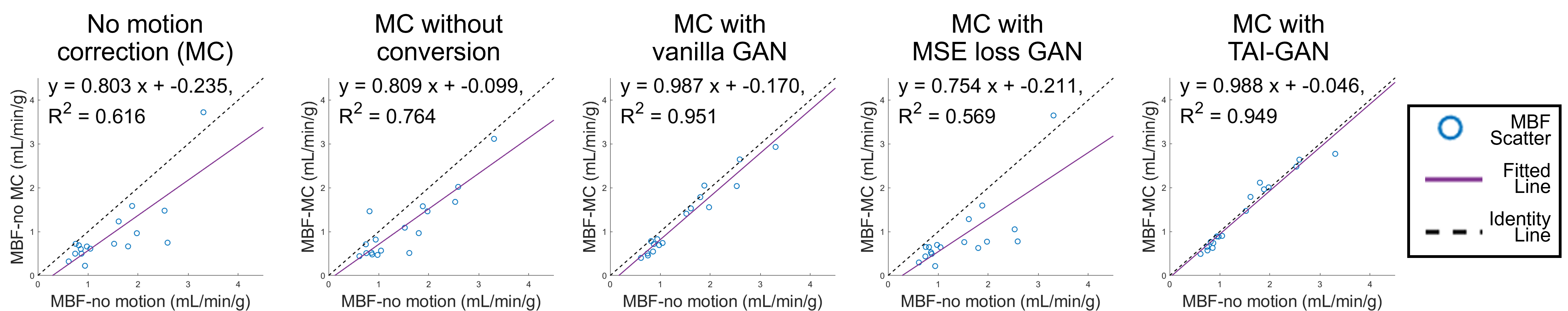}
\caption{Scatter plots of MBF results estimated from motion-free frames vs. no motion correction and motion correction after different conversion methods.}
\label{MBF}
\end{figure}

\begin{table}[t]
\centering
\caption{$K_1$ and MBF quantification results (mean ± standard deviation) with the best results marked \textbf{in bold}.}
\label{tab3}
\resizebox{0.7\textwidth}{!}{
\begin{tabular}{c|c|c|c}
\hline
& \tabincell{c}{Mean $K_1$ percentage\\difference (\%)}& \tabincell{c}{Mean MBF percentage\\difference (\%)} & \tabincell{c}{Mean $K_1$ fitting\\error (×$10^{-5}$) } \\
\hline
\tabincell{c}{Motion-free} & - & - & 3.07 ± 1.85 \\
\hline
\tabincell{c}{With motion} & -25.97 ± 18.05$^{*}$ & -36.99 ± 23.37$^{*}$ & 10.08 ± 8.70$^\dag$\\
\hline
Motion corrected (MC) & -17.76 ± 19.51$^{*}$ & -25.09 ± 31.71$^{*}$ & 22.18 ± 24.24$^\dag$\\
\hline
Vanilla GAN+MC & -11.09 ± 9.79$^{*}$ & -16.93 ± 14.84$^{*}$ & 7.72 ± 6.64$^\dag$\\
\hline
MSE loss GAN+MC & -27.99 ± 18.61$^{*}$ & -39.52 ± 23.89$^{*}$ & 13.05 ± 13.16$^\dag$\\
\hline
TAI-GAN+MC & \textbf{-5.07 $\pm$ 7.68} & \textbf{-7.95 $\pm$ 11.99} & \textbf{3.80 $\pm$ 3.00}\\
\hline
\multicolumn{4}{l}{$^{*}$P < 0.05 between the current method and TAI-GAN (paired two-tailed t-test).}\\

\multicolumn{4}{l}{$^{\dag}$P < 0.05 between the current method and the motion-free result (paired two-tailed t-test).}\\
\end{tabular}
}
\end{table}

Figure \ref{MBF} shows the scatter plots of MBF results estimated from motion-free frames vs. no motion correction and motion correction after different conversion approaches. With simulated motion, the MBF estimates were mostly lower than the ground-truth. The fitted line of motion correction with vanilla GAN was closer to the identity line compared with motion correction without conversion. The fitted line of motion correction with MSE loss GAN was next to that of no motion correction with a slight correction effect. The fitted line of motion correction with the proposed TAI-GAN was the closest to the identity line, suggesting the most improvement in MBF quantification.

Table \ref{tab3} summarizes the bias of $K_1$ and MBF as well as the parametric fitting error. The fitting error of TAI-GAN+MC was the lowest among all the test classes and didn't show a significant difference with the motion-free error (p>0.05). The $K_1$ and MBF percentage differences of TAI-GAN+MC were decreased significantly compared to all the other groups.

\section{Conclusion}
We propose TAI-GAN, a temporally and anatomically informed GAN for early-to-late frame conversion to aid dynamic cardiac PET motion correction. The TAI-GAN can successfully perform early-to-late frame conversion with desired visual results and high quantitative similarity to the real last frames. Frame conversion by TAI-GAN can aid conventional image registration for motion estimation and subsequently achieve accurate motion correction and MBF estimation. Future work includes the evaluation of real patient motion and validation of clinical impact using invasive catheterization as the clinical gold standard.


\clearpage
%
%
%
%
\bibliographystyle{splncs04}
\bibliography{refs.bib}

\begin{thebibliography}{10}
\providecommand{\url}[1]{\texttt{#1}}
\providecommand{\urlprefix}{URL }
\providecommand{\doi}[1]{https://doi.org/#1}

\bibitem{ak2020semantically}
Ak, K.E., Lim, J.H., Tham, J.Y., Kassim, A.A.: Semantically consistent text to
  fashion image synthesis with an enhanced attentional generative adversarial
  network. Pattern Recognition Letters  \textbf{135},  22--29 (2020)

\bibitem{ak2019semantically}
Ak, K.E., Lim, J.H., Tham, J.Y., Kassim, A.: Semantically consistent
  hierarchical text to fashion image synthesis with an enhanced-attentional
  generative adversarial network. In: 2019 IEEE/CVF International Conference on
  Computer Vision Workshop (ICCVW). pp. 3121--3124. IEEE (2019)

\bibitem{burckhardt2009cardiac}
Burckhardt, D.D.: Cardiac positron emission tomography: Overview of myocardial
  perfusion, myocardial blood flow and coronary flow reserve imaging. Mol. Imag
   (2009)

\bibitem{cao2018region}
Cao, X., Yang, J., Gao, Y., Wang, Q., Shen, D.: Region-adaptive deformable
  registration of ct/mri pelvic images via learning-based image synthesis. IEEE
  Transactions on Image Processing  \textbf{27}(7),  3500--3512 (2018)

\bibitem{cciccek20163d}
{\c{C}}i{\c{c}}ek, {\"O}., Abdulkadir, A., Lienkamp, S.S., Brox, T.,
  Ronneberger, O.: 3d u-net: learning dense volumetric segmentation from sparse
  annotation. In: International conference on medical image computing and
  computer-assisted intervention. pp. 424--432. Springer (2016)

\bibitem{dey2021generative}
Dey, N., Ren, M., Dalca, A.V., Gerig, G.: Generative adversarial registration
  for improved conditional deformable templates. In: Proceedings of the
  IEEE/CVF international conference on computer vision. pp. 3929--3941 (2021)

\bibitem{germino2016quantification}
Germino, M., Ropchan, J., Mulnix, T., Fontaine, K., Nabulsi, N., Ackah, E.,
  Feringa, H., Sinusas, A.J., Liu, C., Carson, R.E.: Quantification of
  myocardial blood flow with 82 rb: Validation with 15 o-water using
  time-of-flight and point-spread-function modeling. EJNMMI research
  \textbf{6},  1--12 (2016)

\bibitem{guo2022inter}
Guo, X., Wu, J., Chen, M.K., Liu, Q., Onofrey, J.A., Pucar, D., Pang, Y., Pigg,
  D., Casey, M.E., Dvornek, N.C., et~al.: Inter-pass motion correction for
  whole-body dynamic pet and parametric imaging. IEEE Transactions on Radiation
  and Plasma Medical Sciences  (2022)

\bibitem{guo2022mcp}
Guo, X., Zhou, B., Chen, X., Liu, C., Dvornek, N.C.: Mcp-net: Inter-frame
  motion correction with patlak regularization for whole-body dynamic pet. In:
  Medical Image Computing and Computer Assisted Intervention--MICCAI 2022: 25th
  International Conference, Singapore, September 18--22, 2022, Proceedings,
  Part IV. pp. 163--172. Springer (2022)

\bibitem{guo2021interframe}
Guo, X., Zhou, B., Pigg, D., Spottiswoode, B., Casey, M.E., Liu, C., Dvornek,
  N.C.: Unsupervised inter-frame motion correction for whole-body dynamic pet
  using convolutional long short-term memory in a convolutional neural network.
  Medical Image Analysis  \textbf{80},  102524 (2022).
  \doi{10.1016/j.media.2022.102524}

\bibitem{hochreiter1997long}
Hochreiter, S., Schmidhuber, J.: Long short-term memory. Neural computation
  \textbf{9}(8),  1735--1780 (1997)

\bibitem{hunter2016patient}
Hunter, C.R., Klein, R., Beanlands, R.S., deKemp, R.A.: Patient motion effects
  on the quantification of regional myocardial blood flow with dynamic pet
  imaging. Medical physics  \textbf{43}(4),  1829--1840 (2016)

\bibitem{isola2017image}
Isola, P., Zhu, J.Y., Zhou, T., Efros, A.A.: Image-to-image translation with
  conditional adversarial networks. In: Proceedings of the IEEE conference on
  computer vision and pattern recognition. pp. 1125--1134 (2017)

\bibitem{joshi2011unified}
Joshi, A., Scheinost, D., Okuda, H., Belhachemi, D., Murphy, I., Staib, L.H.,
  Papademetris, X.: Unified framework for development, deployment and robust
  testing of neuroimaging algorithms. Neuroinformatics  \textbf{9}(1),  69--84
  (2011)

\bibitem{liu2019image}
Liu, X., Jiang, D., Wang, M., Song, Z.: Image synthesis-based multi-modal image
  registration framework by using deep fully convolutional networks. Medical \&
  Biological Engineering \& Computing  \textbf{57},  1037--1048 (2019)

\bibitem{lu2020patient}
Lu, Y., Liu, C.: Patient motion correction for dynamic cardiac pet: Current
  status and challenges. Journal of Nuclear Cardiology  \textbf{27},
  1999--2002 (2020)

\bibitem{mao2019bilinear}
Mao, X., Chen, Y., Li, Y., Xiong, T., He, Y., Xue, H.: Bilinear representation
  for language-based image editing using conditional generative adversarial
  networks. In: ICASSP 2019-2019 IEEE International Conference on Acoustics,
  Speech and Signal Processing (ICASSP). pp. 2047--2051. IEEE (2019)

\bibitem{maul2021x}
Maul, J., Said, S., Ruiter, N., Hopp, T.: X-ray synthesis based on triangular
  mesh models using gpu-accelerated ray tracing for multi-modal breast image
  registration. In: Simulation and Synthesis in Medical Imaging: 6th
  International Workshop, SASHIMI 2021, Held in Conjunction with MICCAI 2021,
  Strasbourg, France, September 27, 2021, Proceedings 6. pp. 87--96. Springer
  (2021)

\bibitem{perez2018film}
Perez, E., Strub, F., De~Vries, H., Dumoulin, V., Courville, A.: Film: Visual
  reasoning with a general conditioning layer. In: Proceedings of the AAAI
  Conference on Artificial Intelligence. vol.~32 (2018)

\bibitem{prior2012quantification}
Prior, J.O., Allenbach, G., Valenta, I., Kosinski, M., Burger, C., Verdun,
  F.R., Bischof~Delaloye, A., Kaufmann, P.A.: Quantification of myocardial
  blood flow with 82 rb positron emission tomography: clinical validation with
  15 o-water. European journal of nuclear medicine and molecular imaging
  \textbf{39},  1037--1047 (2012)

\bibitem{rachmadi2019predicting}
Rachmadi, M.F., del C.~Vald{\'e}s-Hern{\'a}ndez, M., Makin, S., Wardlaw, J.M.,
  Komura, T.: Predicting the evolution of white matter hyperintensities in
  brain mri using generative adversarial networks and irregularity map. In:
  Medical Image Computing and Computer Assisted Intervention--MICCAI 2019: 22nd
  International Conference, Shenzhen, China, October 13--17, 2019, Proceedings,
  Part III. pp. 146--154. Springer (2019)

\bibitem{shi2021automatic}
Shi, L., Lu, Y., Dvornek, N., Weyman, C.A., Miller, E.J., Sinusas, A.J., Liu,
  C.: Automatic inter-frame patient motion correction for dynamic cardiac pet
  using deep learning. IEEE Transactions on Medical Imaging  (2021)

\bibitem{shi2019direct}
Shi, L., Lu, Y., Wu, J., Gallezot, J.D., Boutagy, N., Thorn, S., Sinusas, A.J.,
  Carson, R.E., Liu, C.: Direct list mode parametric reconstruction for dynamic
  cardiac spect. IEEE transactions on medical imaging  \textbf{39}(1),
  119--128 (2019)

\bibitem{sundar2021conditional}
Sundar, L.K.S., Iommi, D., Muzik, O., Chalampalakis, Z., Klebermass, E.M.,
  Hienert, M., Rischka, L., Lanzenberger, R., Hahn, A., Pataraia, E., et~al.:
  Conditional generative adversarial networks aided motion correction of
  dynamic 18f-fdg pet brain studies. Journal of Nuclear Medicine
  \textbf{62}(6),  871--879 (2021)

\bibitem{sundar2021data}
Sundar, L.S., Iommi, D., Spencer, B., Wang, Q., Cherry, S., Beyer, T., Badawi,
  R.: Data-driven motion compensation using cgan for total-body [18f] fdg-pet
  imaging (2021)

\bibitem{woo2011automatic}
Woo, J., Tamarappoo, B., Dey, D., Nakazato, R., Le~Meunier, L., Ramesh, A.,
  Lazewatsky, J., Germano, G., Berman, D.S., Slomka, P.J.: Automatic 3d
  registration of dynamic stress and rest 82rb and flurpiridaz f 18 myocardial
  perfusion pet data for patient motion detection and correction. Medical
  physics  \textbf{38}(11),  6313--6326 (2011)

\bibitem{zhou2023fast}
Zhou, B., Tsai, Y.J., Zhang, J., Guo, X., Xie, H., Chen, X., Miao, T., Lu, Y.,
  Duncan, J.S., Liu, C.: Fast-mc-pet: A novel deep learning-aided motion
  correction and reconstruction framework for accelerated pet. In:
  International Conference on Information Processing in Medical Imaging. pp.
  523--535. Springer (2023)

\end{thebibliography}

\clearpage

\title{Supplemental Information}
%
%
\author{Submission ***}
\authorrunning{Submission ***}
%
\institute{Anonymous Organization \\
\email{***@***.***}}
\maketitle              

\renewcommand\thefigure{S\arabic{figure}} 
\renewcommand\thetable{S\arabic{table}} 

\begin{figure}
\centering
\includegraphics[width=\textwidth,keepaspectratio]{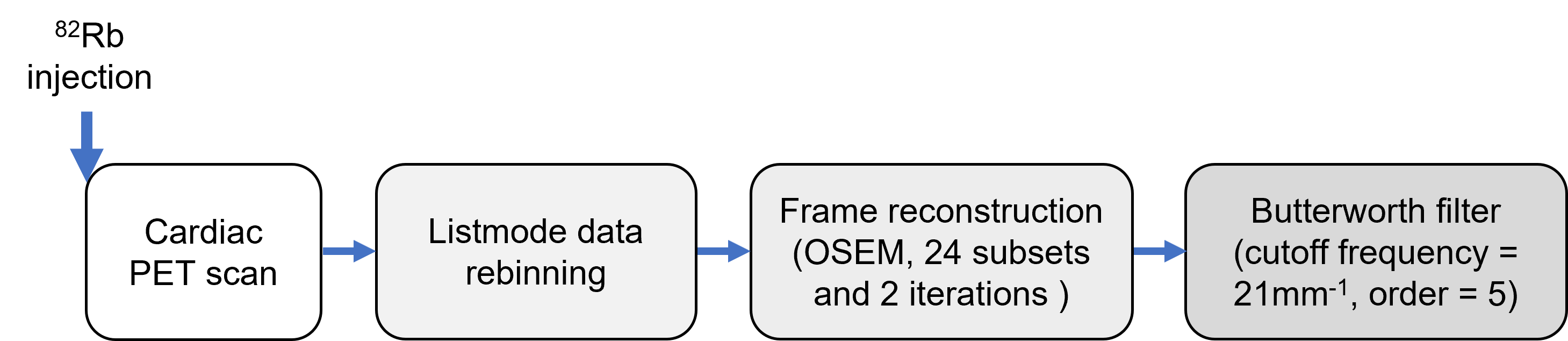}
\caption{The pipeline of $^{82}$Rb cardiac dynamic PET imaging protocol. All the reconstructed dynamic frames have a spatial resolution of 128 × 128 × 47 and a voxel size of 3.125 × 3.125 × 3.270 mm. The corrections of decay, attenuation, scatter, random, prompt-gamma coincidences, detector efficiency, and deadtime were all implemented in the reconstruction process. The image intensities are reconstructed as activity concentration (Bq/mL). } 
\label{imaging_protocol}
\end{figure}

\begin{figure}
\centering
\includegraphics[width=\textwidth,keepaspectratio]{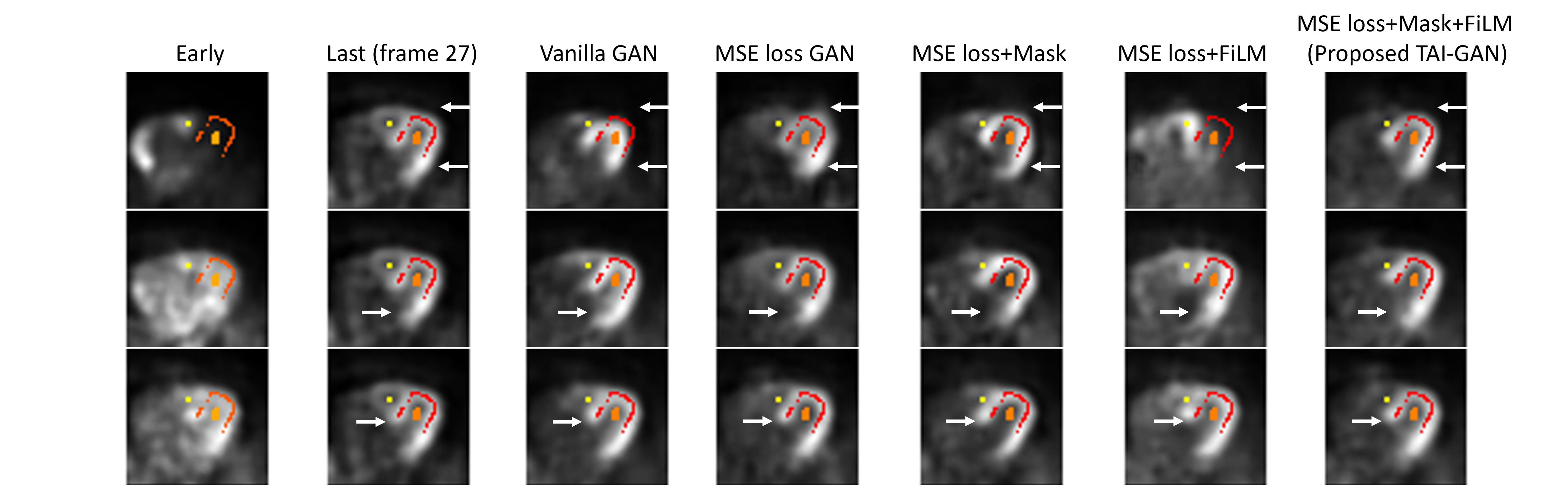}
\caption{Sample results of the preliminary ablation study of the introduced temporal and anatomic information with overlaid segmentations of RVBP (yellow), LVBP (orange), and myocardium (red).}
\label{ablation}
\end{figure}

\begin{figure}
\centering
\includegraphics[width=\textwidth,keepaspectratio]{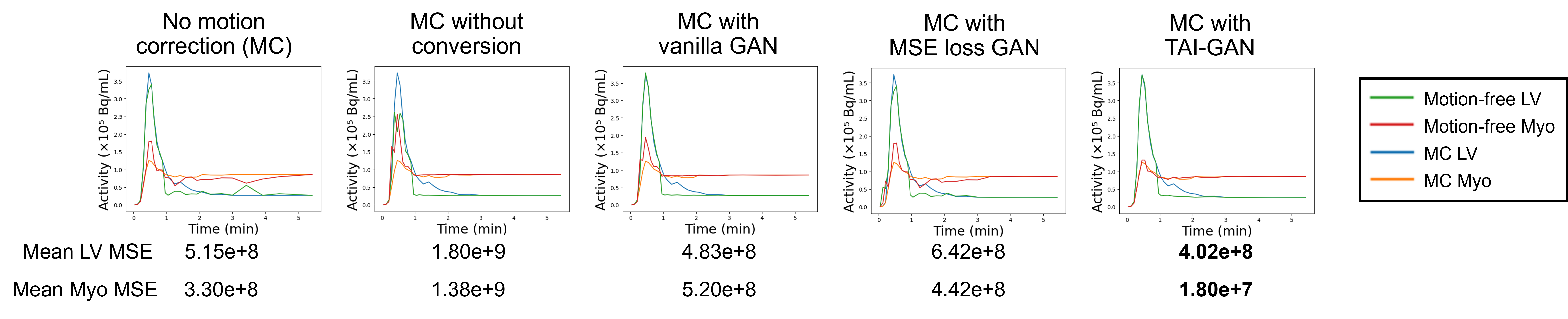}
\caption{A comparison of LVBP and myocardium TACs in each conversion method before and after motion correction. After motion correction with no conversion, the discrepancy was reduced in later frames but drastically increased in the very early blood pool phase frames. The vanilla GAN conversion improved the correction of these frames but remained mismatched. The MSE loss GAN further corrected the blood pool frames but introduced errors in the later frames. The TAI-GAN-aided motion correction achieved the TACs closest to the ground truth with the lowest MSE. 
}
\label{TACs}
\end{figure}

\begin{table}[t]
\centering
\caption{Quantitative image similarity evaluation results (mean ± standard deviation) of the preliminary ablation study of the introduced temporal and anatomic information with the best results marked \textbf{in bold}.}
\label{tabS1}
\resizebox{\textwidth}{!}{
\begin{tabular}{c|c|c|c|c|c|c}
\hline
Test set & Metric & Vanilla GAN & MSE loss & MSE loss+Mask & MSE loss+FiLM & \tabincell{c}{MSE loss+Mask+FiLM \\(Proposed TAI-GAN)}\\
\hline

\multirow{4}{*}{\tabincell{c}{All\\Pre-EQ\\frames}} & SSIM & 0.594 ± 0.068$^{*}$ & 0.637 ± 0.047$^{*}$ & 0.653 ± 0.070$^{*}$ & 0.627 ± 0.068$^{*}$ & \textbf{0.668 ± 0.061}\\
 \cline{2-7}
 & MSE & 0.061 ± 0.023$^{*}$ & 0.038 ± 0.010$^{*}$ & 0.035 ± 0.014 & 0.057 ± 0.026$^{*}$ & \textbf{0.035 ± 0.013}\\
 \cline{2-7}
 & NMAE & 0.069 ± 0.016$^{*}$ & 0.055 ± 0.010$^{*}$ & 0.054 ± 0.012 & 0.063 ± 0.017$^{*}$ & \textbf{0.052 ± 0.012}\\
 \cline{2-7}
 & PSNR & 18.41 ± 1.47$^{*}$ & 20.45 ± 1.22$^{*}$ & 20.86 ± 1.70$^{*}$ & 18.98 ± 2.15$^{*}$ & \textbf{20.89 ± 1.70}\\
 \hline
  \hline

 \multirow{4}{*}{\tabincell{c}{All \\frames}} & SSIM & 0.718 ± 0.082$^{*}$ & 0.722 ± 0.065$^{*}$ & 0.740 ± 0.079$^{*}$ & 0.746 ± 0.087$^{*}$ & \textbf{0.765 ± 0.078}\\
 \cline{2-7}
 & MSE & 0.028 ± 0.019$^{*}$ & 0.022 ± 0.012$^{*}$ & 0.019 ± 0.011 & 0.022 ± 0.018$^{*}$ & \textbf{0.017 ± 0.012}\\
 \cline{2-7}
 & NMAE & 0.048 ± 0.014$^{*}$ & 0.041 ± 0.011$^{*}$ & 0.043 ± 0.011 & 0.043 ± 0.014 & \textbf{0.036 ± 0.012}\\
 \cline{2-7}
 & PSNR & 22.48 ± 2.74$^{*}$ & 23.26 ± 2.19$^{*}$ & 23.91 ± 2.48$^{*}$ & 23.77 ± 3.19$^{*}$ & \textbf{24.65 ± 3.06}\\
 \hline
\multicolumn{6}{l}{{\footnotesize  $^{*}$P < 0.05 between the current class and TAI-GAN (subject-wise paired two-tailed t-test).}}\\ 

\end{tabular}
}
\end{table}

\begin{table}[t]
\centering
\caption{A comparison of the average training time and memory footprint for each frame conversion method.}
\label{tabS2}
\resizebox{\textwidth}{!}{
\begin{tabular}{c|c|c|c|c}
\hline
Metric & Vanilla GAN one-to-one & Vanilla GAN & MSE loss GAN & TAI-GAN (Proposed)\\
\hline
\tabincell{c}{Average training time for\\ all frame conversions (min)} & 780 & 89 & 78 & 98 \\
\hline
Memory footprint (MB) & 2671 & 2671 & 2671 & 3759 \\
 \hline

\end{tabular}
}
\end{table}
\end{document}